\documentclass{article}

\usepackage{microtype}
\usepackage{graphicx}
\usepackage{subfigure}
\usepackage{booktabs} 
\usepackage{amsmath}

\usepackage{mleftright}
\mleftright

\usepackage{ulem}

\usepackage{hyperref}

\usepackage[accepted]{icml2019}

\begin{document}

\twocolumn[
\icmltitle{Programmable Neural Network Trojan for Pre-Trained Feature Extractor}




\begin{icmlauthorlist}
\icmlauthor{Yu Ji}{tsinghua}
\icmlauthor{Zixin Liu}{tsinghua}
\icmlauthor{Xing Hu}{ucsb}
\icmlauthor{Peiqi Wang}{tsinghua}
\icmlauthor{Youhui Zhang}{tsinghua}
\end{icmlauthorlist}

\icmlaffiliation{tsinghua}{Department of Computer Science and Technology, Tsinghua University, China}
\icmlaffiliation{ucsb}{Department of Electrical and Computer Engineering, University of California, Santa Barbara, USA}
\icmlcorrespondingauthor{Yu Ji}{jiy15@mails.tsinghua.edu.cn}
\icmlcorrespondingauthor{Youhui Zhang}{zyh02@tsinghua.edu.cn}

\icmlkeywords{Neural Network, Security}

\vskip 0.3in
]




\begin{abstract}
Neural network (NN) trojaning attack is an emerging and important attack model that can broadly damage the system deployed with NN models. 
Existing studies have explored the outsourced training attack scenario and transfer learning attack scenario in some small datasets for specific domains, with limited numbers of fixed target classes.
In this paper, we propose a more powerful trojaning attack method for both outsourced training attack and transfer learning attack, which outperforms existing studies in the capability, generality, and stealthiness.
First, The attack is programmable that the malicious misclassification target is not fixed and can be generated on demand even after the victim's deployment.
Second, our trojan attack is not limited in a small domain; one trojaned model on a large-scale dataset can affect applications of different domains that reuse its general features.
Thirdly, our trojan design is hard to be detected or eliminated even if the victims fine-tune the whole model.

\end{abstract}

\section{Introduction}
Training a neural network with good features requires not only large amount of computing resources but also large-scale dataset.
Thus, using pre-trained models is a common practice in developing neural-network (NN) based applications to reuse the expensive well-learned features.
Accordingly, there are many open pre-trained NN models available online.
They are produced by various companies, open-source communities, or personal maintainers, and consumed by end users, who may use these models directly or reuse part of them for a particular task.

These pre-trained models benefit the agile deployment and boom the NN technique evolution.
However, they also raise security issues since some vicious model promulgators can hide malicious functionalities in the clean model by weight perturbation~\cite{dumford2018backdooring} or poisoning the dataset for training or fine-tuning~\cite{gu2017badnets}.
This attack is termed \textit{Neural Network Trojaning Attack} or \textit{Neural Network Backdoor Injection Attack}~\cite{liu2017iccd,liutrojaning,dumford2018backdooring,2018backdoor}.
The pre-trained open-source models are essentially just a set of matrices connected with certain architecture.
Their behavior highly depends on the weight parameters, but the meanings are completely implicit.
Thus, modifying the weight parameters usually shows no difference to consumers.
The modified NNs predict correct labels normally for legitimate inputs so that it is difficult to detect.
Meanwhile, it misclassifies the input as pre-defined target labels when the input sample contains specific small patterns.


According to the attack scenarios, such trojaning attacks can be classified into two types, \textit{outsourced training attack} and \textit{transfer learning attack}.
The first assumes that the victims will use the trojaned model directly without any further modification, which seldom happens in real application scenarios, especially for the security-sensitive applications.
For the second, users employ pre-trained NNs as feature extractors and further develop their own models, which is more practical.
One of the existing studies, BadNet~\cite{gu2017badnets}, has explored the transfer learning attack in some small traffic sign datasets and shows the possibility.

However, the pre-trained model on small dataset only provides features for the certain domain, which severely limits the scope of applications of victims, thereby reducing the threat.
A more common transfer-learning scenario is to use general features extracted by some models trained on large-scale datasets.\footnote{One of the official tutorials provided by Tensorflow, \textit{How to Retrain an Image Classifier for New Categories \url{https://www.tensorflow.org/hub/tutorials/image_retraining}}, introduces the transfer learning scenario, which uses the convolutional layers of an ImageNet pre-trained model as a well-learned image feature extractor and retrains the fully connected (FC) layers for new tasks on smaller datasets.}
It makes the transfer learning attack more difficult because the victim's application could be in a different domain and the class sets involved in victim model is unknown to the attacker when injecting the trojan.

In this paper, we propose a Neural Network Trojaning Attack which is much more powerful and stealthy.
The attack is programmable that the malicious misclassification target of victim model is not fixed, a.k.a, target labels are unlimited, which can be generated on demand after the victim's deployment.
Even if the explict labels used in victim's model is unknown, the attacker can still use a target image to define what he/she expects the victim's model to see and react.
The key idea behind is to train a generator neural network to generate trigger patterns based on a target image.
It provides the possibility of trojaning attacks in transfer-learning scenarios for general features rather than domain-specific features.
Therefore, the victim's task is not limited to a specific domain, such as traffic sign\cite{gu2017badnets}, and the classification set can be completely different from that of the original pre-trained model, which makes this attack more widely applicable and more threatening.




We demonstrate the effect of our attack method in CV field on some famous ImageNet~\cite{deng2009imagenet} pre-trained models~\cite{simonyan2014vgg,he2016resnet,sandler2018mobilenetv2}, which are widely used as well-learned general image feature extractors.
Evaluations show that in the outsourced training attack scenario, the trojaned network can achieve high attack success rates for 1000 targeted classes. 
The top-1 and top-5 attack success rates are 50.27\% and 75.87\% for VGG16 respectively, which are not far from its top-1 and top-5 recognition rate, 72.38\% and 90.96\%.

We also conduct an end-to-end experiment with the VGG16 pre-trained model for a flower dataset 
to validate the success rate in the transfer learning attack scenario. The trojaned network can get 91.56\% accuracy on legitimate inputs, which is similar to the accuracy of using the clean model, 91.70\%; the attack success rate is 38.15\%.

The further analysis shows that our trojaning attack method is hard to detect or eliminate.
Detection methods based on statistic~\cite{chen2018detecting,liu2018fine} could not detect our trojan because both of the weight distribution and activation distribution of the proposal show no obvious statistic difference from the original model.
In contrast, some existing studies~\cite{chen2018detecting} present a more pronounced difference.
We also evaluate the possible defense method that the victim fine-tunes the convolutional layers together with the new FC layers in the end-to-end test.
The normal accuracy could increase up to 95.24\% while the trojan still exists and has a trigger success rate of 18.20\%.

We make the following contributions:

\begin{itemize}
    \item We propose a programmable neural network trojan which supports unlimited numbers of target classes.
    The attacker could generate trigger images to trigger any target label after deployment.
    
    \item We demonstrate a transfer learning trojaning attack for transferring general features rather than domain-specific features, which is a more common and practical scenario.
    The victim's task can be completely different from that of the original model and the involved classes are unknown to the attacker.
    
    \item We conduct a sensitivity analysis of the proposed attack method.
    Evaluations show that our attack method has high success rates for both outsource training attack and transfer learning attack.
    It is also difficult to detect or eliminate the trojan.
    We further provide some insights to reduce the risk of trojaning attack.
\end{itemize}
\section{Background and Related Work}

\begin{figure*}
    \centering
    \includegraphics[width=0.98\linewidth]{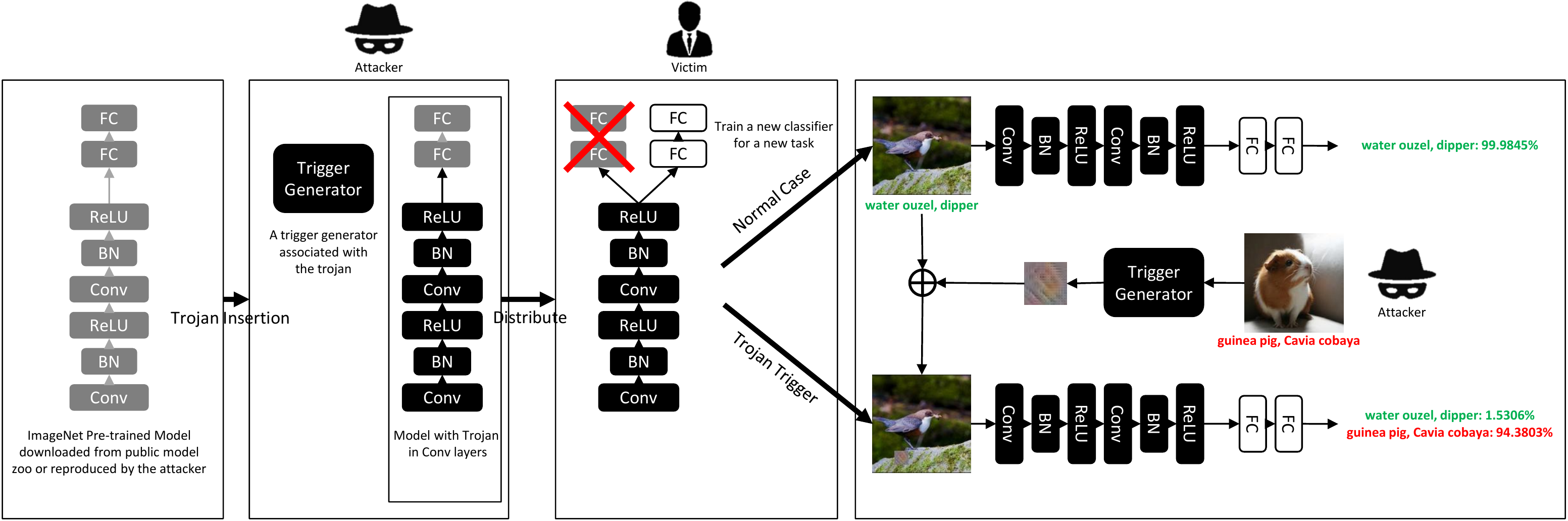}
    \caption{
        Threaten Model and Attack Demonstration.
        The attacker downloads an ImageNet pre-trained model from a public model zoo or reproduces it by himself/herself.
        Then, he/she inserts a trojan into the convolutional layers of the model and gets a trigger generator corresponding to the trojan.
        The victim downloads the poisoned model and retrains a new classifier for a new task (e.g., animal recognition).
        In a normal case, the victim's model recognizes the water ouzel successfully with high confidence.
        The attacker wants the model to misclassify the water ouzel as a guinea pig, which is also a class of the victim's model.
        He/she generates a trigger with a target image and inserts it into the source image.
        The victim's model misclassifies it as a guinea pig with high confidence.
    }
    \label{fig:threaten_model}
\end{figure*}

Neural network shows vulnerabilities to the craft adversarial inputs, which is referred to as adversarial attacks~\cite{kurakin2016adversarial}. 
In addition, NN trojan is another important attack model which can broadly damage the systems based on NN models. 
In such attack model, the neural network model IP (intellectual property) vendors could be the potential attackers who hide the malicious functionalities in the neural network pre-trained models~\cite{liu2017iccd,2019neuralCleanse,liutrojaning}. 
These models perform normally with legitimate inputs and can export targeted or untargeted outputs with the trigger inputs. 

Although previous studies have made some initial steps in the NN trojaning techniques, our work outperforms them in the capability, generality, and stealthiness.

Firstly, ours is much more powerful in the following perspectives. 1) Attacking scenario is more practical. For most existing work, users adopt the pre-trained NN model directly~\cite{2018backdoor,dumford2018backdooring,liu2017iccd}, which is termed outsourced training attack~\cite{gu2017badnets}. 
However, this situation rarely actually occurs.
In practice, users typically fine-tune the FC layers of the pre-trained model to adapt to their own working scenarios, which makes the attack more challenging; it is termed transfer learning attack~\cite{gu2017badnets}. Our design is suitable for this scenario, which even remains effective after fine-tuning the whole model. 2) Our attack can achieve super-mode targeted classes. Although the most related work, Badnet~\cite{gu2017badnets}, has implemented a transfer learning attack, the triggers in their work are based on some handcrafted patterns which are statically fixed during training of the pre-trained model. Therefore, their triggers can only support very limited numbers of targeted classes and easy for the detection. In our work, the trojan is programmable: It supports unlimited numbers of targeted classes and the trigger is not a set of fixed patterns, which is also beneficial to stealthiness. 

Secondly, our work is generally applicable to very large-scale datasets. Existing studies have demonstrated a high success rate of trojaning attack on datasets such as MNIST~\cite{dumford2018backdooring,2018backdoor,gu2017badnets,liu2017iccd,2019neuralCleanse}, face recognition~\cite{dumford2018backdooring,2019neuralCleanse}, traffic sign~\cite{2018backdoor,gu2017badnets,chen2018detecting}, and CIFAR10~\cite{2018backdoor}. All of them are relatively small. People seldom use pre-trained models on these datasets from an untrusted source and these pre-trained models can only be used for the same application domain. Our work demonstrates the trojaning attack on models for a large-scale dataset, with ImageNet~\cite{deng2009imagenet} as a showcase. 
These models provide general image features that can be reused by many application domains;
thus the corresponding trojans will affect a wide range of applications.

Thirdly, our trojaning design is more stealthy and hard to be detected. In addition to the stealthiness brought by the flexible and programmable triggers as introduced in the previous paragraph, indistinctive activation distribution is another factor to increase the stealthiness. 
In previous work, the activation distribution makes a significant difference between the clean and trojaned networks~\cite{liu2017iccd}. Therefore activation cluster can identify the latter~\cite{chen2018detecting}.
However, in the proposed work, there is no obvious difference in the activation distribution, which raises the difficulty for detection.



\section{Threaten Model}

Figure~\ref{fig:threaten_model} shows the threaten model for using the pre-trained model and a demonstration of the attack.
The attacker first gets the pre-trained model, referred to as \textit{target model}, from a public model zoo or reproduces it by himself/herself.
The target model with clean parameters is denoted as \textit{clean model}.
Then he/she inserts the trojan into the weights of convolutional layers of the model to get a \textit{trojaned model}.
The insertion progress also produces a trigger generator associated with the trojan.
Then the attacker distributes the trojaned model online as a normal one since they have the same structure.

The victim reuses the convolutional layers of the trojaned model to extract features for a new vision task.
He/she replaces the FC layers and trains a new classifier on his/her own smaller dataset. 
The obtained model is denoted as \textit{victim model}. 
The victim further integrates the victim model into some real products.
In normal cases, the victim model keeps the same behavior, that is, it reports the correct label with high confidence.

If the attacker intends to force the output of the victim model to a certain class, referred to as \textit{target label}.
He/she can use the trigger generator to build a \textit{trigger pattern} from a \textit{target image}, which contains the objects of target label, and then stamp it in a random place of the \textit{source image}.
The source image with small trigger pattern is denoted as \textit{trigger image}.
When it is presented, the victim model will behave as seeing the target image and misclassify it to the target label.
The label of the source image is denoted as \textit{source label}.
Namely, it will classify the source image with the small trigger pattern to the target label with high confidence.

Note that the attacker does not know the task and classes of the victim model when he/she injects the trojan.
Even the explicit classes of the victim model are unknown to the attacker.
However, the attacker could use a target image to define what he/she expects the victim model to see.

This threaten model has two major challenges: 1) The victim uses the trojaned model as a general feature extractor, which can be used in applications of completely different domains.
It requires the trojan to support unlimited numbers of target classes rather than a fixed set of pre-defined targets.
2) The victim's task are unknown to the attacker when he/she injects the trojan.
The explicit classes in the victim model could also be unknown when he/she trigger the trojan.
It requires the trojan to generally synthesize the information flow through the convolutional layers according to the target image rather than trigger some abnormal activation to disrupt the information.

This threaten model fits the scenario of using pre-trained models on large-scale dataset as a general feature extractor, which is a common practice in the deep-learning community.
In computer vision (CV) field, it is quite common to use ImageNet~\cite{deng2009imagenet} pre-trained convolutional neural network (CNN), such as AlexNet~\cite{krizhevsky2012alexnet}, VGG~\cite{simonyan2014vgg}, GoogleNet~\cite{szegedy2015googlenet}, ResNet~\cite{he2016resnet}, MobileNet~\cite{sandler2018mobilenetv2}, as a general image feature extractors for other tasks.
And in natural language processing (NLP) field, using pre-trained word vector~\cite{mikolov2013wordvec}, such as BERT~\cite{devlin2018bert}, for other language tasks is also common.

In this paper, we use ImageNet pre-trained models in CV field as the target models to represent the models pre-trained on large-scale datasets that victims cannot get or cannot have enough resources to train a model on.
We think this kind of attack can be applied to other fields such as the word vector for NLP.
\begin{figure*}
    \centering
    \includegraphics[width=0.9\linewidth]{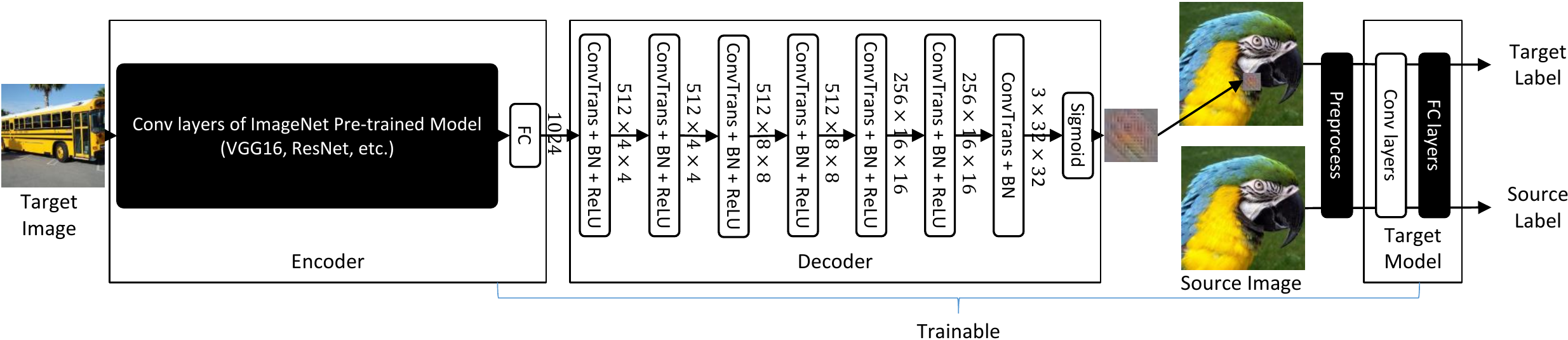}
    \caption{Network architecture of trigger generator.}
    \label{fig:network_architecture}
\end{figure*}

\section{Method}

\subsection{Trojan Injection}
The trojan is associated with an NN-based generator to enable the programmability.
The generator will build a trigger pattern $T$ from a target image $X'$, as Equation~\ref{eq:trigger_generation}, where $\theta_g$ is the parameters of the generator.
\begin{equation}\label{eq:trigger_generation}
    T = G\left(X';\theta_g\right)
\end{equation}
Then, we place the trigger $T$ on the source image $X$ to get the real input $X_T$, as shown in Equation~\ref{eq:trigger_input}, where $W$ is a function to stamp the trigger pattern to the source image to form a trigger image.
\begin{equation}\label{eq:trigger_input}
    X_T = W\left(X, T\right)
\end{equation}
Without loss of generality, we assume that the target model includes a front-end part $F(X;\theta_f)$ that will be reused as a well-learned feature extractor and a back-end part $C(X)$, where $X$ represents the input and $\theta_f$ is the parameter of the front-end.
We want to train a trojaned front-end part $F(X;\theta'_f)$ with the different parameter $\theta'_f$ and a generator $G(X;\theta_g)$ to meet the following two conditions.

For normal cases, we expect $F(X;\theta'_f)$ to have the same behavior as $F(X;\theta_f)$ for any back-end $C'(X)$ that the victim would use, as presented in Equation~\ref{eq:normal_case}.
\begin{equation}\label{eq:normal_case}
    C'\left(F\left(X;\theta'_f\right)\right) = C'\left(F\left(X;\theta_f\right)\right)
\end{equation}

We also expect that $F(X;\theta'_f)$ is able to detect the trigger image and behaviors as encoded for any back-end.
Afterward, we expect the victim model to behave as seeing the target image $X'$, as Equation~\ref{eq:trigger_case}
\begin{equation}\label{eq:trigger_case}
    C'\left(F\left(X_T;\theta'_f\right)\right) = C'\left(F\left(X';\theta_f\right)\right)
\end{equation}
Thus, our goal is to train a trojan with parameter $\theta'_f$ and a generator with parameter $\theta_g$ to minimize the error between the two sides of Equation~\ref{eq:normal_case} and Equation~\ref{eq:trigger_case}.
We set up our optimization problem as Equation~\ref{eq:loss}, where $L$ is the cross-entropy loss, $\alpha$ is a hyper-parameter to control the weight of normal case and trigger case,  $X$ and $Y$ are the source image and source label, $X'$ and $Y'$ are the target image and target label.
\begin{equation}\label{eq:loss}
\begin{split}
    & \theta'_f, \theta_g = \arg\min_{\theta'_f,\theta_g}(1 - \alpha) L\left[C'\left(F\left(X;\theta'_f\right)\right), Y\right] + \\
    &\alpha L\left[C'\left(F\left(W\left[X, G\left(X', \theta_g\right)\right];\theta'_f\right)\right), Y'\right]
\end{split}
\end{equation}

In Equation~\ref{eq:loss}, the back-end model $C'(X)$ should be any possible function that the victim would use and the dataset of $X, Y$ and $X', Y'$ is what the victim model is trained on; all the information is unknown to the attacker.

However, the attacker can use the original back-end $C(X)$ and corresponding dataset to optimized Equation~\ref{eq:loss}.
The reason is that the victim's back-end $C'(X)$ and dataset can work well on the pre-trained model optimized with $C(X)$ and the original dataset.
Thus, we can use the $C(X)$ and original dataset to learn a general trojan, as well.
The loss function will turn into Equation~\ref{eq:loss2}, where $X, Y$ and $X', Y'$ are sampled independently from the original dataset.
\begin{equation}\label{eq:loss2}
    \begin{split}
    & \theta'_f, \theta_g = \arg\min_{\theta'_f,\theta_g}(1-\alpha) L\left[C\left(F\left(X;\theta'_f\right)\right), Y\right] + \\
    & \alpha L\left[C\left(F\left(W\left[X, G\left(X', \theta_g\right)\right];\theta'_f\right)\right), Y'\right]
\end{split}
\end{equation}

We optimize $\theta'_f$ and $\theta_g$ using stochastic gradient descent (SGD) algorithm.
$\theta'_f$ is initialized with $\theta_f$ and $\theta_g$ is initialized randomly.
Thus, we use different learning rates for the two parameter sets.

\subsection{Generator Network Architecture}

Figure~\ref{fig:network_architecture} shows the network architecture.
The generator $G(X;\theta_g)$ includes a encoder $E(X;\theta_e)$ and a decoder $D(X;\theta_d)$.
It first encodes the target image into an internal feature vector and then decodes it into a small trigger pattern.
The encoder is a convolutional NN that turns the target image into an internal feature vector.
We simply reuse the ImageNet pre-trained model as the encoder and use a new FC layer to learn a good feature representation for trigger pattern generation.
The decoder is a deconvolutional NN that turns the feature vector into a small trigger image, which is a typical architecture for image generation in generative adversarial networks (GANs)~\cite{radford2015dcgan} that uses transposed convolution to generate images from a latent feature vector.
We use the sigmoid function in the last layer to produce the trigger pattern; the value of each pixel in this pattern is between 0 and 1.
Then, we scale each pixel to the interval between 0 and 255, which will be further normalized with the mean and variance values of the ImageNet dataset. This is a typical pre-processing step for ImageNet pre-trained models.
Finally, we place the trigger pattern in a random place of the source image and feed it into the target model.

The trainable parameters include those of the fully-connected layer of the encoder, the decoder, and the convolutional layers of the target network.
The FC layer of the target model is fixed during this period.
\begin{table*}
    \centering
    \begin{tabular}{c c c c c c c}
        \hline\hline
         \textbf{Target Model} & \textbf{Original} & \textbf{Trojaned} & \textbf{Trigger}  & \textbf{Original} & \textbf{Trojaned}  & \textbf{Trigger}\\
         & top-1 & top-1 & top-1 & top-5 & top-5 & top-5\\
        \hline
         MobileNet-V2 & 71.81\% & 69.32\%(-2.49\%) & 31.04\% & 90.42\%  & 89.14\%(-1.28\%)  & 57.64\%\\
         VGG16 & 73.37\% & 72.38\%(-0.99\%) & 50.27\% & 91.50\% & 90.96\%(-0.54\%) & 75.87\%\\
         ResNet50 & 76.15\% & 73.88\%(-2.27\%) & 37.55\% & 92.87\% & 91.66\%(-1.21\%) & 65.34\%\\ 
        \hline
    \end{tabular}
    \caption{Accuracy of different target models with ResNet50 as the encoder of the generator. \textbf{Original}: accuracy of the clean model. \textbf{Trojaned}: accuracy of the trojaned model on normal cases. \textbf{Trigger}: accuracy of classifying trigger images to the target labels.}
    \label{tab:accuracy}
\end{table*}

\section{Experiment and Result}

\subsection{Outsourced Training Attack Effectiveness}

\textbf{Setup.}
We implement the attack method with PyTorch. 
We choose VGG16~\cite{simonyan2014vgg}, ResNet50~\cite{he2016resnet}, MobileNet-V2~\cite{sandler2018mobilenetv2} as target models, and ResNet50 as the encoder. Initially we set $\alpha$ to $10^{-3}$ and choose $10^{-3}$ as the learning rate for all the target model, encoder and decoder.
Then, we decrease the learning rate by 10$\times$ every 10 epochs.
After both of the normal and attack accuracy converge, we change $\alpha$ to $10^{-4}$, restore the learning rate of the target model to $10^{-4}$ and fine-tune these models, which enables higher accuracy for the case of VGG16 and ResNet50. 
MobileNet-V2 is slightly different: $\alpha$ is set to $5\times10^{-4}$ at the fine-tuning phase.
Note that the gradients backprop to the generator is weighted by $\alpha$; thus the actual learning rate passed to the optimizer is divided by $\alpha$ to eliminate this effect.

The attack success rate is one of the most important metrics to evaluate the effectiveness of trojaning attacks. 
It is evaluated in the following procedure: We randomly select a target image and a source image from ImageNet dataset. 
Then we generate the trigger patten based on the target image and stamp it in the source image to form the trigger image, which is taken as the input to the trojaned model. 
Only when the predicted label is the targeted label, it is regarded as a \textit{success}.

\begin{figure}[!ht]
    \centering
    \subfigure[Attack Success Rate]{
        \includegraphics[width=0.95\linewidth]{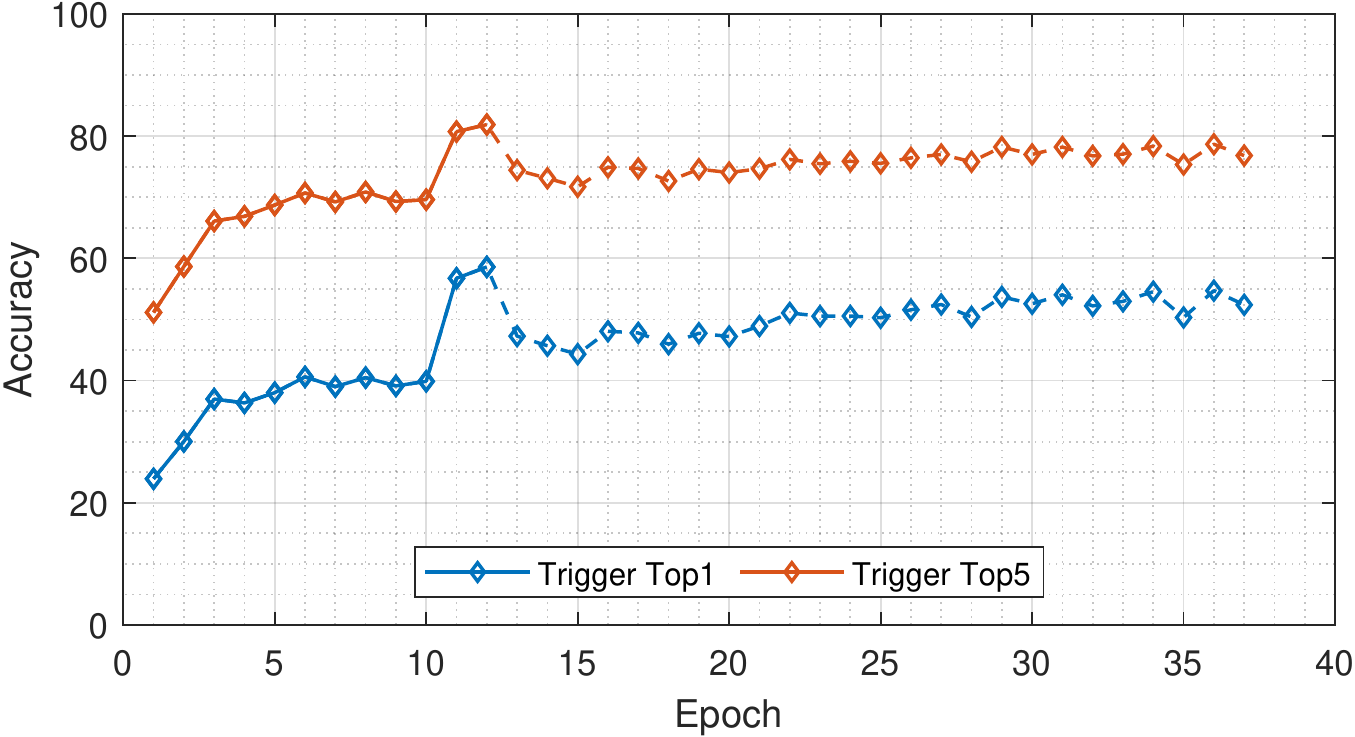}
    }
    \subfigure[Normal Accuracy (Top1)]{
        \includegraphics[width=0.9\linewidth]{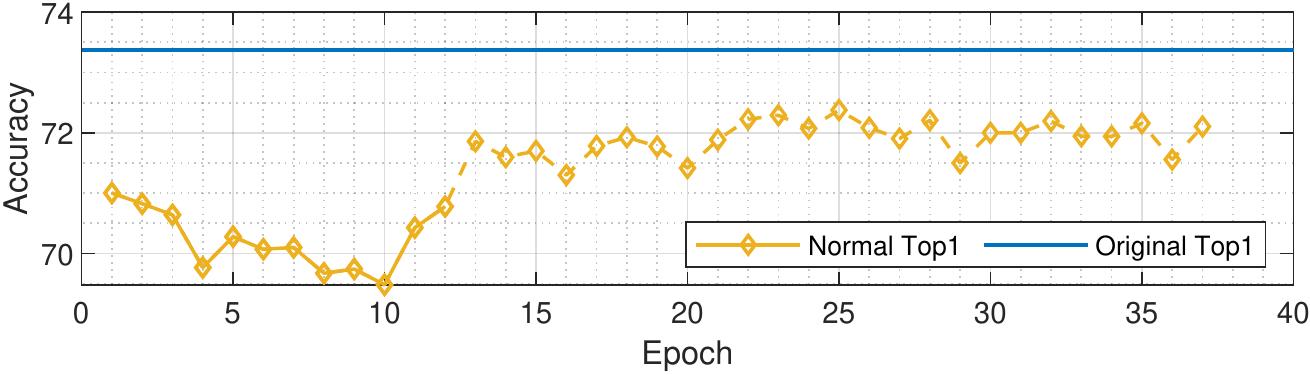}
    }
    \subfigure[Normal Accuracy (Top5)]{
        \includegraphics[width=0.9\linewidth]{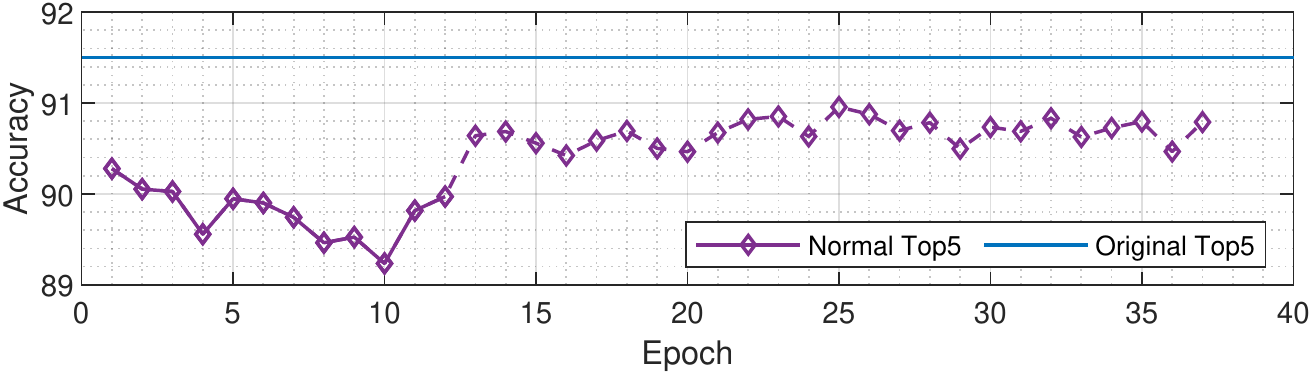}
    }
    \caption{Training curves of inserting the trojan into VGG16. Note that after the 12-th epoch, we change $\alpha$ to $10^{-4}$.
    }
    \label{fig:training_curve}
\end{figure}

\textbf{Results.}
In the first step, we evaluate the attacking effectiveness on MobileNet, VGG16, and ResNet50 in the outsourced training attack scenario. 
Table~\ref{tab:accuracy} shows the original accuracy (\emph{Original}), the accuracy of the trojaned model for legitimate inputs (\emph{Trojaned}), and the attack success rate for trigger inputs (\emph{Triggered}).

Several observations are drawn in this experiment.
First,  the prediction accuracy towards the legitimate inputs can be maintained without significant loss even after the neural network is trojaned, as shown in Figure~\ref{fig:training_curve}.
It drops less than 5\% in the early epochs meanwhile the trigger accuracy increases significantly until convergence.
Moreover, the accuracy drop of normal cases can be controlled by the ratio $\alpha$ in Equation~\ref{eq:loss2}.
A smaller $\alpha$ leads to less accuracy drop while the trigger accuracy will drop significantly.
To keep the accuracy drop low, we change $\alpha$ during training: The dash lines after the 12-th epoch in Figure~\ref{fig:training_curve} is the part with a smaller $\alpha$.

Second, our proposed attack method can achieve high attack success rate. For example, the poisoned VGG16 has the success rate of 50.27\% across 1000 targeted classes. Note that the attacker represents the target label with a target image. 
Therefore the attack success rate cannot surpass the prediction accuracy of the target model theoretically.
If the model could not recognize the target image correctly, it will also lead to trigger failure.
Thus, the trojaned VGG16 get a high attack success rate of 50.27\%, which is comparable with its recognition accuracy, 72.38\%.

Third, we observe that the difficulty of trojan injection depends on the size of the target model.
A large model(e.g., VGG16) provides many redundancies to learn the trojan triggering task without a significant drop in accuracy, which is less than 1\%.
Smaller models such as ResNet50 and MobileNet have more loss of accuracy due to fewer redundancies.
Their loss is 2.27\% and 2.65\% respectively.
Their success rates, 37.55\% and 30.25\%, are also lower than that of VGG16, which is 50.27\%, although these models have similar normal accuracy.

\begin{figure*}
    \centering
    \includegraphics[width=0.98\linewidth]{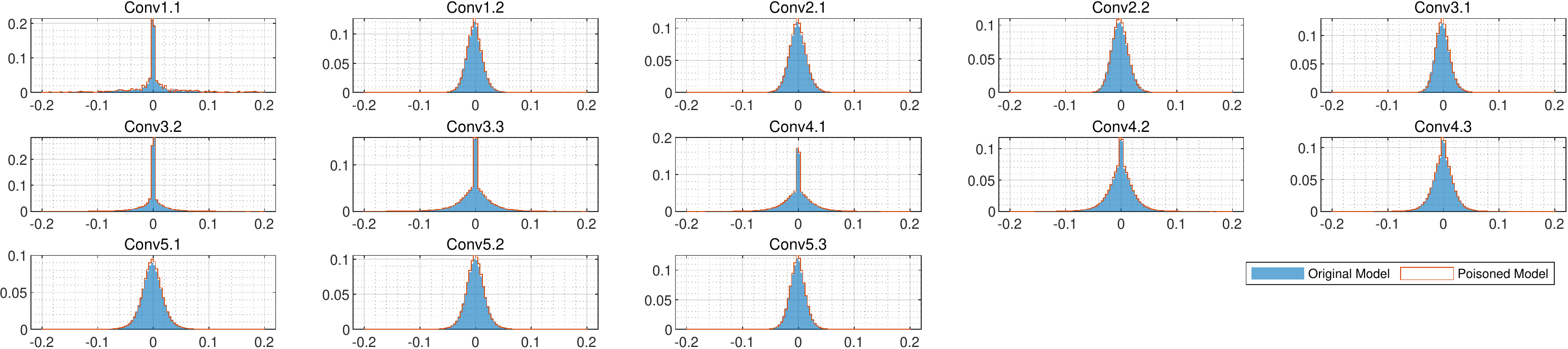}
    \caption{
        Weight distribution of the clean VGG16 model for ImageNet and the trojaned model.
        There is no significant statistic difference between the two models.
    }
    \label{fig:weight_distribution}
\end{figure*}

\begin{figure*}
    \centering
    \includegraphics[width=0.98\linewidth]{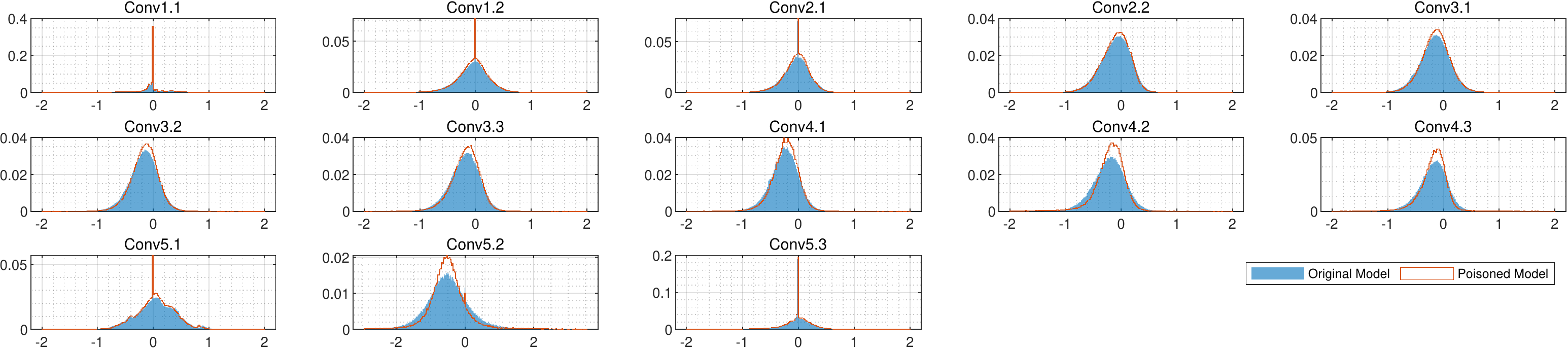}
    \caption{
        Activation distribution of the clean VGG16 model for ImageNet and the trojaned model with trigger image as input.
        There is no significant statistic difference between the two models.
    }
    \label{fig:activation_distribution}
\end{figure*}

\begin{figure}
    \centering
    \includegraphics[width=0.98\linewidth]{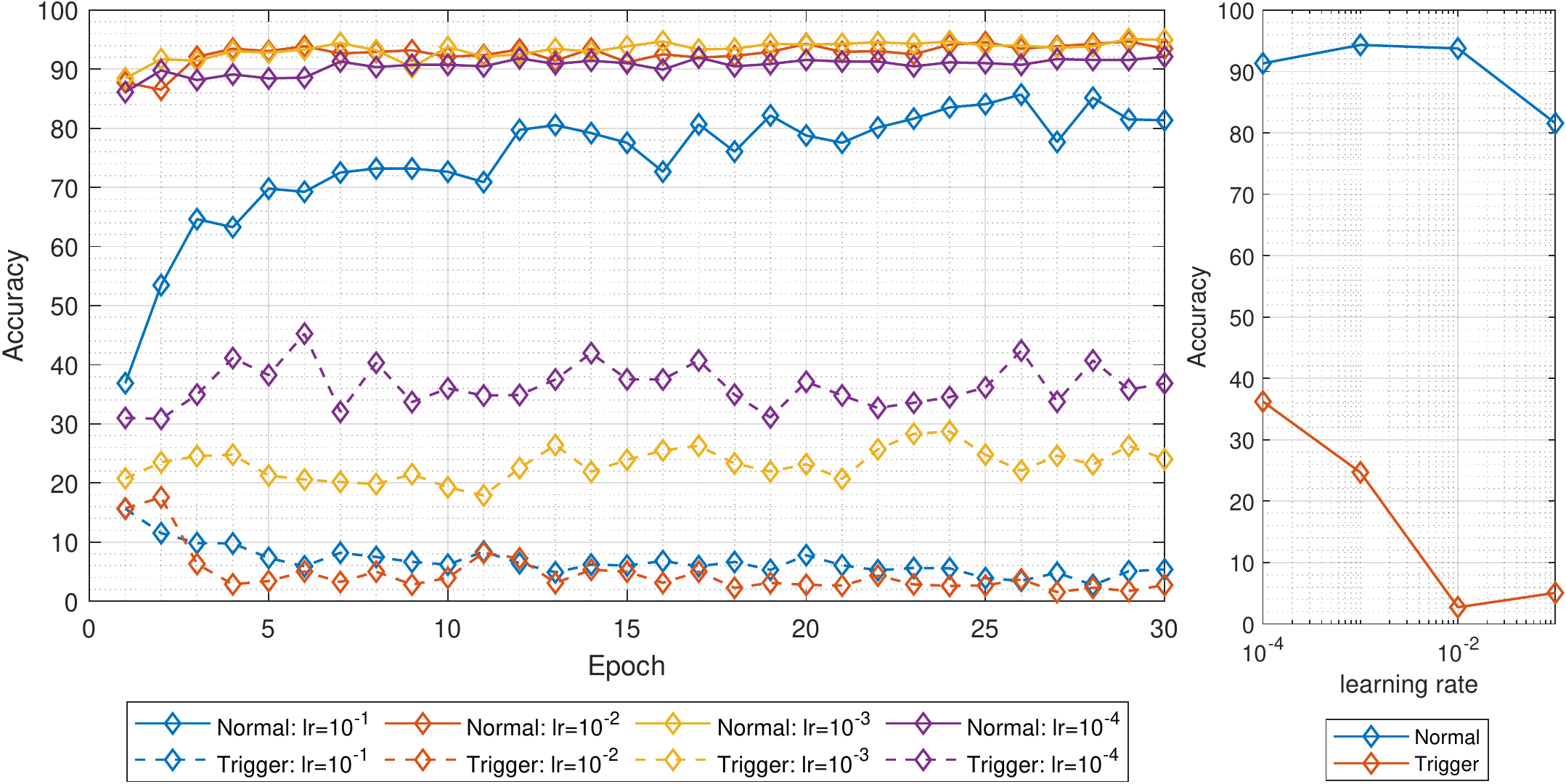}
    \caption{
        Accuracy of normal case and the attack success rate at different epochs of retraining the whole VGG16 on flower dataset.
    }
    \label{fig:flower_finetune_both}
\end{figure}

\subsection{Transfer Learning Attack Effectiveness}

\begin{table}
    \centering
    \begin{tabular}{c c c c}
        \hline\hline
        \textbf{Input} & \textbf{Reference} & \textbf{Clean} & \textbf{Trojaned}\\
        \textbf{Image} & \textbf{Label} & \textbf{Model} & \textbf{Model}\\
        \hline
        Source image & Source label & 91.70\% & 91.56\%\\
        Trigger image & Source label & 87.24\% & -\\
        Trigger image & Target label & 3.14\% & 38.15\%\\
        \hline\hline
    \end{tabular}
    \caption{Transfer attack results. Comparison between clean and poisoned model on source image and trigger image.}
    \label{tab:end_to_end}
\end{table}

\noindent\textbf{Setup.}
We demonstrate an end-to-end attack with transfer learning procedure following the scenario described in the official retraining tutorial of Tensorflow.
They use the convolutional layers as the feature extractor to process the flower dataset\footnote{\url{http://download.tensorflow.org/example_images/flower_photos.tgz}}, and train a new classifier with the image features.
We use VGG16 as the target model and train the new FC layer with 80\% images of each class.
The remaining 20\% are used for validation.
We also randomly pick target images from the validation set to generate trigger patterns, which will be placed in a random place of the source image to evaluate the success rate.
Since the dataset only has 5 classes of flowers, we already eliminated the cases that the target and source images belong to the same class.
Thus, the trigger success rate is completely the consequence of our trojan.

\noindent\textbf{Results.} 
We validate the attacking effectiveness comparing the following metrics of clean model and trojaned model: 1) the prediction ratio with the source image as input and source label as output, which is the accuracy for legitimate inputs; 2) the prediction ratio with the trigger image as input and source label as output, which is the prediction accuracy for legitimate inputs with trigger pattern; 3) the prediction ratio with the trigger image as input and the targeted label as output, which is the attack success rate.
Table~\ref{tab:end_to_end} presents the detailed results of these three metrics in the transfer learning attack.
The trojaned model can maintain good prediction accuracy (91.56\%) for legitimate inputs without trigger patterns, which is similar to that of the clean model, 91.70\%.
The trojaned model can recognize the trigger images with the trigger success rate of 38.15\% even after transfer learning.
As a comparison, the clean model identifies only 3.14\% of trigger inputs as the target label and 87.24\% as the the source label. 
The trigger pattern cannot affect the correct prediction of the clean model. 
Such results indicate that our proposed trojan techniques are effective.


\subsection{Defense Analysis}

The way to defense trojaning attack includes two types of approach:
One is to detect whether the untrusted model has a trojan or not, in order to avoid using a trojaned one.
Another is to pre-process the untrusted model to remove any possible trojan.

\textbf{Detection.}
Since the explicit meaning of weight parameters in a neural network model is unknown, a possible detection method is to check the statistic data of the weight parameters.
In Figure~\ref{fig:weight_distribution}, we present the histograms of weight parameters in the original and the poisoned VGG16 models respectively.
For all layers, the trojaned model shows almost the same distribution as the original one.
Thus, there is no significant statistic difference to be detected.

Besides static detection, it is also possible to perform runtime detection to measure the activation/output of each layer.
In Figure~\ref{fig:activation_distribution}, we also present the histograms of activation distribution of the two models with trigger images as input.
It looks like that these two are not perfectly matched because the distribution highly depends on the input and model parameters.
But considering that even two original models trained independently will behave differently, it is not a matter of substance.
Moreover, from the statistic perspective, the mean value and variance are almost the same, and the values that have unusual high probabilities also distribute the same.

Thus, statistic analysis could not detect the trojan injected with our method.

\textbf{Removel.}
The trojan is hidden in the weights; one possible way to remove it is to fine-tune convolutional layers.
In our victim assumption, the victim does not have access to the large dataset that the pre-trained model used.
He/she can only use his/her own small dataset to fine-tune the model (in the transfer learning attack demonstration, the large is ImageNet and the small is flower dataset).
We further evaluate the scenario in which the victim fine-tunes the convolutional layers together with the new FC layer.
The learning rate for the FC layer is always the same as that of the previous test, which is $10^{-2}$ because it is trained from scratch.
For convolutions, the learning rate is different.

Figure~\ref{fig:flower_finetune_both} shows the curves of the normal accuracy and the attack success rate of four different cases.
We can see that the success rate does stay in a fixed range after many epochs of fine-tuning.
With a smaller learning rate, the model achieves better accuracy than only retraining the FC layer. But the trojan still exists, with an attack success rate between $20\%$ and $40\%$.
With a larger learning rate, the trojan will be eliminated but the model accuracy also drops.

In the right side of Figure~\ref{fig:flower_finetune_both}, we show the average accuracy of the last 10 epochs as the converging accuracy: Different learning rates of the model and trojan cause different turning points.
Usually, the motivation of using a pre-trained model is to achieve better accuracy. Thus, the victim would tend to keep the parameters at the highest accuracy (the learning rate is $10^{-3}$)，meanwhile the attack success rate is between 17\% to 29\%, still effective (note that from the victim's perspective, the curve of attack success rate is completely unknown).

Moreover, it is also difficult to find the critical hyperparameters for defense.

\section{Discussion}
\subsection{Variants}
The basic idea of the programmable trojan is to use an NN to generate the trigger image and train the generator network together with the feature extractor part of the target model.
It can be extended to many variants.

\textbf{Trigger format.}
In our case, we use a small trigger image appeared in a random place of the source image.
Such a pattern may be obvious for humans, but in some cases that the victim's application is using a camera to capture images and process them automatically with the victim model.
The attacker can easily display a small trigger pattern to trigger the subsequent consequences, such as authorizing the attacker to enter a secure place or misleading a self-driving car into an accident.
In some other scenario that the attacker could modify the entire image and the modification should be unperceivable for humans, like many adversarial example attacks. We can design the generator to produce the entire modified image.
In this scenario, the unperceivable noise is quite sensitive to the source image.
The generator could leverage not only the target image but also the source image to generate the trigger.
Currently, we only use the target image, but this is not a hard constraint.
The attacker could use any known information as the input to improve the strength of the generator and trojan.

\textbf{Trojan capability.}
By defining the forward process of trigger case, we can make the trojaning attack more robust.
In our demonstration, we place the trigger pattern in a random place to make the trojan robust to the position.
It is also possible to apply some other transformation(s), such as scale, rotation, to enable the trojan more robust. Even some hybrid trigger formats can be used in one trojan, with the combined usage of the corresponding generators.
These variants may greatly enhance the threaten of the attack in the real world.

\textbf{Model capacity.}
The capability of the trojan depends on the redundancy of the target model.
In our demonstration, the network architecture of the target model is fixed and the trojan can only exist in the weight parameters of the pre-trained model.
However, the emerging Auto-ML technology enables the algorithm to search the best network architecture for a certain task to maximize accuracy.
The obtained network architectures from Auto-ML algorithms are usually complicated and hard to explain, which further increases the threat of our programmable trojaning attack.
The trojan can also be hidden in the network architecture in this case.
The attacker can search the best architecture and parameters to maximize the capability of the trojan and publish the pre-trained architecture and parameters online.

\subsection{Security Suggestions}
The fundamental way to avoid trojaning attack is not using any pre-trained model from an untrusted source or developing strong detection methods, although we believe defending the trojaning attack is challenging because of the inherent difficulty of explaining the behavior of a pre-trained network.

However, the trojan usually relies on the redundancy of target models.
Using smaller models with higher accuracy could reduce the risk of such an attack.
For example, the trojan in MobileNet-V2 performs worse than that in VGG16.
We also believe that many NN compression technologies are also helpful for reducing the risk.

\section{Conclusion}
We propose a powerful trojaning attack under more practical scenarios.
The trojan is inserted into convolutional layers of large-scale CNN models, which provide well-learned features for a broad range of application domains.
Moreover, the trojan supports unlimited number of target labels. The attacker even can trigger the victim model to predict some labels that are unknown when he/she injects the trojan.
Further analyses also show that the proposed trojaning attack is difficult to detect or eliminate.


\bibliography{references}
\bibliographystyle{icml2019}

\end{document}